# A Novel Magic LSB Substitution Method (M-LSB-SM) using Multi-Level Encryption and Achromatic Component of an Image

Khan Muhammad[1], Muhammad Sajjad[1], Irfan Mehmood[1], Seungmin Rho[2], Sung Wook Baik[1*]
Tel: +82-02-3408-3797, Fax: +82-02-3408-4339
khan.muhammad.icp@gmail.com, sajjad@sju.ac.kr, irfanmehmood@sju.ac.kr,
smrho@sungkyul.ac.kr, sbaik@sejong.ac.kr
[1]*College of Electronics and Information Engineering, Sejong University, Seoul, Korea*
[2]*Department of Multimedia, Sungkyul University, Anyang, Korea*

**Abstract**
Image Steganography is a thriving research area of information security where secret data is embedded in images to hide its existence while getting the minimum possible statistical detectability. This paper proposes a novel magic least significant bit substitution method (M-LSB-SM) for RGB images. The proposed method is based on the achromatic component (I-plane) of the hue-saturation-intensity (HSI) color model and multi-level encryption (MLE) in the spatial domain. The input image is transposed and converted into an HSI color space. The I-plane is divided into four sub-images of equal size, rotating each sub-image with a different angle using a secret key. The secret information is divided into four blocks, which are then encrypted using an MLE algorithm (MLEA). Each sub-block of the message is embedded into one of the rotated sub-images based on a specific pattern using magic LSB substitution. Experimental results validate that the proposed method not only enhances the visual quality of stego images but also provides good imperceptibility and multiple security levels as compared to several existing prominent methods.

## 1. Introduction

Steganography is a special branch of information hiding where a secret message is embedded in a cover image based on a shared stego key, resulting in a stego image [1-3]. In contrast to steganography, steganalysis aims to detect or extract the hidden data in those stego images. The steganographic algorithm is considered to be broken if an attacker can decide whether or not a given image is a stego image, based on steganalysis with a higher probability of detection instead of just random guessing [4]. Steganography requires a carrier object, secret data and an embedding algorithm. It also requires an encryption algorithm and a secret key in some cases, increasing the security levels of steganography. Applications of steganography includes secure transmission of top-secret documents between national and international governments, captioning, tamper-proofing, securing online banking, voting systems, and time-stamping [5, 6]. Watermarking and cryptography are two closely related areas to steganography. The main theme of steganography and cryptography is same, i.e., to obscure the secret information, but the corresponding techniques used in both areas are different. The procedure of steganography and watermarking are similar, carrying different purposes. Steganography deals with the embedding of secret data while watermarking is concerned with copyright protection of digital data [7].

---

[*] Corresponding author: E-mail: sbaik@sejong,ac.kr

Steganographic methods are broadly classified into spatial domain and transform domain methods. In the spatial domain, the gray levels of the original carrier image are directly modified for encoding the secret data. These techniques employ a high payload but are vulnerable to image processing manipulations and statistical attacks such as image cropping, image compressing, noise attacks and chi-square attack. Some examples of spatial domain techniques include LSB[8-11], gray-level modification method[12], edges based embedding techniques[4, 13-16], pixel indicator techniques (PIT)[17, 18], pixel value differencing techniques[19, 20], pixel pair matching method[21], and tri-way pixel value differencing method[22]. In transform domain, the image is converted from the spatial domain to the transform domain and the image coefficients are modified to hide secret information. These techniques have a lower payload but they are more robust against statistical attacks. Some examples of transform domain techniques are the discrete wavelength transform technique[23], discrete Fourier transform technique[24], discrete cosine transform techniques[25, 26], and contourlet transform technique[27].

The simplest and most basic spatial domain steganographic method is LSB substitution, which hides secret data inside a cover image. With this method, the least significant bits of the carrier image pixels are replaced with the secret data bits. Payload capacity of the LSB method can be increased if more than 1 LSBs are used for message embedding, but it makes noticeable changes in the carrier image. Wang et.al,[28] presented a genetic algorithm based on an LSB substitution scheme for improving the stego image quality. The Wang et.al, approach requires more processing time, which is its major shortcoming. To reduce the complexity of the Wang et.al, scheme, Chang et.al,[29] proposed a fast algorithm based on LSB and dynamic programming. Lou and Liu [30] presented an LSB based technique that is capable of hiding various sizes of secret information and is resilient against cover carrier attacks. A novel approach is presented by Lin and Thien [31] based on the LSB method using modulus functions with the same aim of improving the quality of the stego images. Chang and Cheng [6] demonstrated a pixel adjustment based approach for obtaining better quality of stego images. Lin and Tsai [32] nominated a new scheme for addressing the problem of image authentication and enhancing security by making use of steganographic methods and image sharing concepts. Wu et.al,[33] proposed an efficient scheme by combining the LSB method and pixel value differencing method with the goals of attaining a high payload and better quality stego images. The LSB based methods are quite straight forward but it is easy to detect the existence of data embedded via these methods using different steganalysis systems including chi-squared attack[34], sample pair analysis[35], regular-singular (RS) group analysis[6], and structural based steganalysis framework[36].

LSB matching (LSB-M) is another improved version of the LSB approach, which randomly adds +1 or -1 to a given pixel if the message bit is not same as the LSB of that pixel[4]. LSB-M reduces the asymmetry produced by the simple LSB method and is not detectable by steganalysis algorithms that detect data hidden through LSB approaches. To detect the M-LSB based embedded data in stego images, some other steganalysis systems[37, 38] have been proposed.

The LSB methods and LSB-M use the host image's pixels independently. To solve this problem, an improved version of LSB-M is proposed in [11] known as LSB-M revisited (LSB-MR). LSB-MR embeds two secret bits at a time in a pair of pixels. The 1$^{st}$ secret bit is embedded in the 1$^{st}$ pixel and 2$^{nd}$ secret bit is hidden using the relationship between the pixels in that particular pair.

This minimizes the modification rate of the host image in bits per pixel (bpp) from 0.5 to 0.375 with the same capacity as compared to LSB and LSB-M. Furthermore, LSB-MR also reduces the asymmetry caused by the LSB method and makes the extraction of hidden data difficult.

The LSB based approaches described so far embed the secret messages in carrier image pixels regardless of whether a pixel is located at edge area or smooth area. Tsai and Wu [35] proposed a high imperceptible steganographic technique based on the idea that an edge area pixel can carry more secret bits as compared to smooth area pixels. They embed data in image pixels by noting the difference between two consecutive pixels. A larger difference indicates that the current two pixels lie at edge area and are capable of carrying more secret bits. On the other hand, a smaller difference between two consecutive image's pixels, determines that the two pixels are located on a smooth area and a small amount of secret bits can be embedded inside these pixels. Using this concept as a base, a number of edge based techniques have been proposed in the literature[4, 15, 16] [39] [40]. The proposed techniques achieve a high payload and better quality of stego images as compared to LSB based techniques, but security is still a major problem in these approaches as the data is in plain form.

This paper proposes a novel approach for steganography to overcome the limitations of some existing steganographic methods in terms of security and imperceptibility. The main contributions of this paper are:
  i. The achromatic component (I-plane) of an image in an HSI color model is used for embedding instead of an RGB color model to increase the security of the proposed method and reduce the extra computational overhead.
  ii. Secret information is encrypted using MLEA before it is embedded in the carrier image pixels which adds one more level of security to the said technique.
  iii. The secret information and I-plane are divided into four sub-blocks and each of the message blocks is embedded into a specific image block using a new improved version of LSB method ("Magic LSB") which further makes the data extraction more challenging.

The rest of the paper is structured as follows. Section 2 discusses some existing classic and recent steganographic methods in the literature whose defects led us towards the current proposed work. The proposed work is discussed in section 3 which is followed by experimental results and performance analysis in section 4. Section 5 summarizes the concluding remarks of the paper and suggests future directions.

## 2. Literature Review
Digital steganography is a blooming research area that uses digital images, videos, network protocols and audio for information concealment. From the last decade, several approaches for digital steganography have been proposed in the spatial domain. These approaches are based on LSB substitution, edge based embedding and pixel indicator based embedding. In this section, we present a brief overview of the basic LSB method and discuss some other existing state-of-the-art techniques within each category that are related to the proposed method. At the end of this section, we present some strategies to cope up with the limitations of the methods mentioned.

## 2.1 Basic Idea of LSB methods

The basic idea of the LSB method is to replace the least significant bits of the host image with the bits of secret data. To briefly describe this basic idea of a classical LSB substitution scheme, consider *I* as a host 8-bit image having *n* pixels such that $I=I_0I_1...I_{n-1}$ where $I_j$ is a gray level of *I* for *j*=0, 1, 2….*n-1*. Suppose *M* is a secret message such that $M=M_0, M_1....M_{n-1}$ with $M_j$ a *k*-bit string of message *M* for *j*=0, 1….*n-1*. The pixel $I_j$ is divided into two sub-sections in order to hide a bit $M_j$ in the carrier pixel $I_j$. The two sub-sections are $LSB_j$ and $MSB_j$ with $I_j=MSB_j \| LSB_j$ and $LSB_j$ is replaced with $M_j$ for *j*=0, 1….*n-1*. The stego image *S* with pixels $S=S_0, S_1....S_{n-1}$ is obtained after message hiding such that $S_j \in S$ with *j*=0, 1….*n-1*.

Now consider an image *I* with eight (8) pixels $\{I_1-I_8\}$ and secret character using binary representation as follows:

| $I_1$=10001101 | $I_2$=10000010 | $I_3$=01110110 | $I_4$=01100001 |
|---|---|---|---|
| $I_5$=00101000 | $I_6$=10000100 | $I_7$=01001011 | $I_8$=01110111 |
| *Secret character:* **B**→ 01000010 ||||

After replacing the LSB's of these pixels with the bits of secret character *"B"*, the pixels changes to $\{S_1-S_8\}$ as follows:

| $S_1$=1000110**0** | $S_2$=1000001**1** | $S_3$=01110110 | $S_4$=0110000**0** |
|---|---|---|---|
| $S_5$=00101000 | $S_6$=10000100 | $S_7$=01001011 | $S_8$=0111011**0** |

By noticing the resultant pixels, it can be observed that only half of the pixels change ($S_1, S_2, S_4,$ and $S_8$). This classical LSB method increases or decreases the pixel value by 1 or leaves it unchanged depending upon the LSBs of the image pixels and the bits of secret information. In Figure 1(a), a host image of Lena with dimension (256×256 pixels) is given. After replacing one LSB (*k=1*) of each pixel with the bits stream of message ("*Welcome to the great seat of learning; Islamia College Peshawar, Khyber Pakhtunkhwa, Pakistan*"), the resultant stego image is obtained as shown in Figure 1(b).

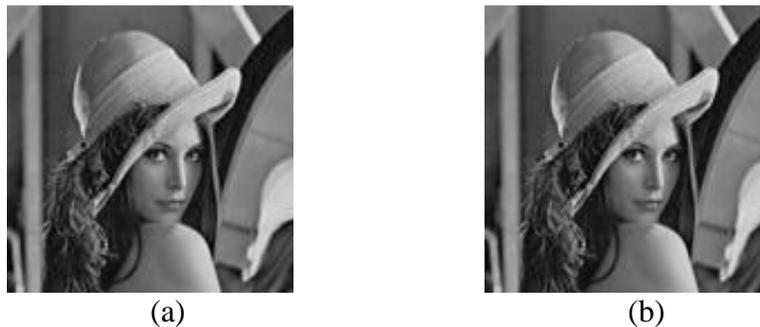

(a)            (b)

Figure 1: An example of LSB substitution method. a): Lena cover image and b) corresponding stego image with *k*=1.

Figure 1(a) and 1(b) clearly show that the asymmetry artifacts caused in the stego image is almost negligible and cannot be observed by human visual system (HVS). Payload can be increased by increasing the value of *k* i.e. to replace more than 1 LSBs of the host image pixels but it causes obvious distortion in the stego image. In Figure 2, different stego images of Lena are shown by changing its various planes i.e. *k=2*, *k=3*, *k=4* and *k=5*.

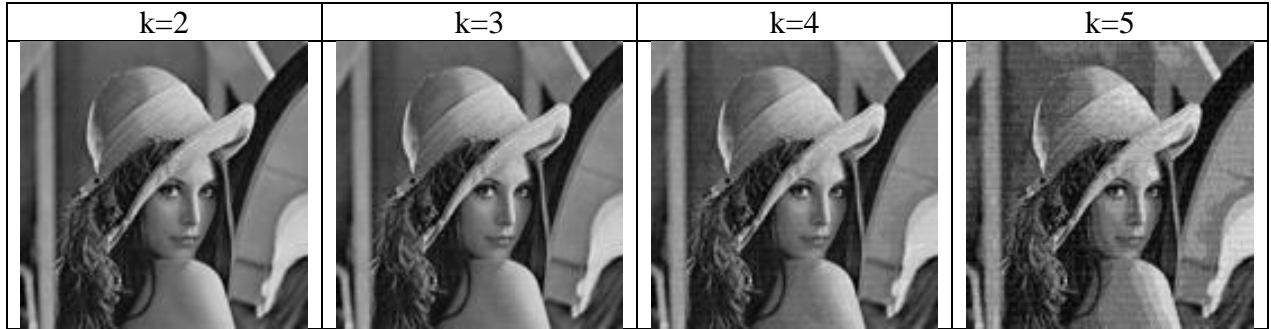

Figure 2: Degradation in the quality of Lena stego image by hiding data in different image planes.

LSB-M slightly modifies the image pixels by adding $\pm 1$ randomly to the gray levels of the host image when the secret bit does not match the LSB of a given pixel, keeping the values of pixels in the range 0-255. The extraction process of LSB and LSB-M is same i.e. to generate a traversing path using a shared secret key and extracting the LSB of every pixel to get the actual embedded bits. LSB-MR[11] uses a pair of pixels ($P_i$, $P_{i+1}$) as a unit of embedding which is modified into ($P'_i$, $P'_{i+1}$) such that it satisfies the given criteria.

$$\left\{ LSB(P'_i) = S_i \qquad LSB\left(\left\lfloor\frac{P'_i}{2}\right\rfloor + P'_{i+1}\right) = S_{i+1} \right\} \qquad (1)$$

Here, $P_i$ and $P_{i+1}$ show the embedding unit and $S_i$ and $S_{i+1}$ represent the two secret bits. Using this relationship, the LSB and LSB-M like asymmetry artifacts are not produced in stego images. Furthermore, LSB-MR reduces the rate of modification in terms of pixels in contrast to LSB and LSB-M method. In the extraction process, a traversing path is first generated based on a shared stego key and a pseudo random number generator and then two bits are extracted from each of the embedding units.

**2.2 Cyclic LSB based Approaches**
The LSB based approaches result in stego images of good quality but that can be easily compromised and hacked by attackers as these techniques are quite straightforward. To increase the security and scatter the message in the whole host image, Bailey and Curran [41] proposed the stego color cycle (SCC) method for color images. SCC hides data in different channels of the cover image, allowing dispersion of data throughout the entire image. The mechanism of this approach is cyclic in nature. i.e., the first secret bit is hidden in pixel1's red channel, the second secret bit is hidden in the green channel of pixel2, and the third secret bit is hidden in the blue channel of pixel3, and so on. The major limitation of the SCC method is that the secret information is embedded in the cover image pixels in a fixed cyclic and systematic way. So an attacker can easily discover this technique if he/she successfully extracts data from a few pixels. A modified version of SCC is proposed in [42] using randomization which provides more security as compared to the SCC technique but still the technique is straight forward and extracting data from a few pixels can enable the attacker to extract the hidden data.

## 2.3 Pixel Indicator based Methods

The LSB and cyclic LSB based techniques result in better quality of stego images but these techniques possess lower payloads i.e. 1bpp. To increase the payload, Parvez et.al,[43] proposed the idea of the pixel indicator technique (PIT), which logically divides the three channels of an RGB image into indicator channel and data channels. The indicator channel decides the data channel for data hiding, which continuously changes according to a fixed sequence, allowing better security. The data is embedded in the host image based on the information given in Table 1.

Table 1: Indicator based data hiding[43]

| 1st and 2nd LSB of indicator channel | Data Channel1 | Data Channel2 |
|---|---|---|
| 00 | Nothing to hide | Nothing to hide |
| 01 | Nothing to hide | Replace 2 LSBs of this channel |
| 10 | Replace 2 LSBs of this channel | Nothing to hide |
| 11 | Replace 2 LSBs of this channel | Replace 2 LSBs of this channel |

The PIT method gives better results in terms of payload and security, minimizing the stego key overhead. The capacity of PIT is dependent on the indicator channel and cover image, which can lead to lower payload in some cases. Moreover, it uses a fixed number of bits per channel, causing noticeable distortion in the stego image. Adnan[18] proposed another method to solve these problems by hiding data in channels based on its intensity. The proposed method increases the security of [43] by introducing the stego key for channel selection. Parvez et.al,[44] further increased the security of [18] based on partition schemes. In addition, data is distributed in the cover image using statistical methods. Amirtharajan et.al,[45] proposed a color guided based data hiding method which further improves the security of Pervez et.al method[44]. Swain and Lenka[46] proposed a novel method to further improve the payload of all mentioned PIT based approaches. Amirtharajan et.al,[47] presented a novel scheme based on statistical theory by embedding variable amount of bits in image pixels, for further improving the payload. The security of [47] is enhanced using stego key and randomization by authors in[48].

## 2.4 Edges based Data Hiding Methods

The LSB, cyclic LSB and PIT based methods directly embed data in the pixels of the host image without taking into consideration that a pixel is located at smooth area or edge area. Edge area pixels can accommodate more secret bits as compared to smooth areas and are less detectable by the HVS. Keeping this in mind, Tsai and Wu [35] proposed the first edge based steganographic technique, which increased the payload of the LSB and cyclic LSB methods. Chen et.al, [15] proposed a new approach using a hybrid edge detector that combines the canny and fuzzy edge detectors, increasing the payload of [35]. Lue et.al[4] combined the edges based data hiding method with LSB-MR[11] which resulted in better quality of stego images and a larger payload. To further increase the payload of [15], A. Ioannidou et.al [16] proposed a novel edges based technique for color images whose payload is three times more than the existing methods. Grover et.al, [39] proposed a new method by hiding three bits in edge pixels and two bits in smooth pixels, increasing its payload. Furthermore, the proposed scheme divides the data into two blocks and traverses the pixels starting from the center of the host image, which further increases the security. The quality of stego images in existing edges based hiding methods is fixed. The authors in [40] resolved this issue by proposing a novel method in which the quality of stego images is tunable.

The techniques discussed so far embed secret data directly in the host image without encryption, which makes it easy to extract if the encoding algorithm is compromised by the attacker. Furthermore, some of the mentioned existing methods result in stego images of low quality due to which they can be easily detected by the HVS. In this paper, we propose a novel and secure scheme which overcomes the limitations of some mentioned state-of-the-art methods by M-LSB-SM. A malicious user cannot extract the actual secret message even if the embedding algorithm is known because data is divided into four blocks and is encrypted using MLEA. An attacker has to pass through the following barriers in order to achieve the actual hidden contents of data.

 i. The secret key for rotating the sub-images of the I-plane.
 ii. The detail information about MLEA.
 iii. The steganographic algorithm applied for information concealment.
 iv. Have knowledge about the fact that image has been transposed and achromatic component of HSI color model have been used instead of RGB for encoding of data.
 v. The information that which message block is embedded in which image block.

## 3. The Proposed Scheme

In this section, the proposed method is presented in detail. First, some terminologies related to the proposed method are briefly described in Table 2. Then, we present some mathematical notations and diagrams to briefly introduce the proposed method. Next, MLEA is described in Pseudo code form, followed by embedding algorithm with a suitable example. Finally, the extraction method is briefly discussed by mentioning its major steps.

Table 2: Summary of terminologies and symbols used in the proposed M-LSB-SM scheme

| Terminology/Symbol | Description |
|---|---|
| Cover Image ($I^C$) | The input image in which secret information will be embedded |
| Stego Image ($I^S$) | The image containing the secret information |
| Transposed Image ($I^T$) | The image rotated at 90° |
| HSI Image ($I^{HSI}$) | The image which is converted from RGB color space to HSI color space |
| MGM | Magic Matrix (A special type of matrix in MATLAB) |
| M | M is the secret message that is to be embedded in cover image ($I^C$) |
| MLEA | Multi-Level Encryption Algorithm |
| $B_1, B_2, B_3, B_4$ | The encrypted message sub-blocks returned by MLEA |
| $I^{plane}$ | The achromatic component of $I^{HSI}$ |
| $S^{plane}$ | The chromatic component (Saturation) of $I^{HSI}$ |
| $H^{plane}$ | The chromatic component (Hue) of $I^{HSI}$ |
| $I_{c1}, I_{c2}, I_{c3}, I_{c4}$ | Rotated cover sub-images of $I^{plane}$ |
| $I_{s1}, I_{s2}, I_{s3}, I_{s4}$ | Stego sub-images containing sub-message blocks $B_1, B_2, B_3,$ and $B_4$ |
| $K^{key}$ | The stego/secret key that is used in MLEA and rotations of sub-images |
| KB | The array containing the binary bits of $K^{key}$ |
| Magic LSB | A novel data hiding steganographic method |
| Cipher | The message which is encrypted using MLEA |

Suppose $I^C$ is the carrier image and can be transposed using the function given in equation 2. The existing color space of the transposed image $I^T$ is then converted to HSI color space using equation 3, getting the image $I^{HSI}$ as an output. HSI color space plays an important role in message concealment because changing the I-plane does not affect the other planes of the image unlike RGB in which all the three planes are strongly co-related with each other. Furthermore, processing an image in HSI color space is relatively more cost effective[5].

Suppose $M$ denotes the secret message that is to be embedded into the carrier image $I^C$, $K^{key}$ shows the secret key and $I^S$ represents the stego image containing secret information. Six functions have been used in the proposed process of embedding as shown in equations 2-7.

$$I^T = \text{transpose}(I^C) \tag{2}$$

$$I^{HSI} = \text{RGB2HSI}(I^T) \tag{3}$$

$$[B_1, B_2, B_3, B_4] = \text{MLEA}(M, K^{key}) \tag{4}$$

$$[I_{c1}, I_{c2}, I_{c3}, I_{c4}] = \text{ImageSubDivision}(I^{plane}, K^{key}) \tag{5}$$

$$[I_{s1}, I_{s2}, I_{s3}, I_{s4}] = \text{MagicLSB}([I_{c1}, I_{c2}, I_{c3}, I_{c4}], [B_1, B_2, B_3, B_4]) \tag{6}$$

$$I^S = \text{ReconstructStego}([I_{s1}, I_{s2}, I_{s3}, I_{s4}], K^{key}, H^{plane}, S^{plane}) \tag{7}$$

The message $M$ is encrypted using the *MLEA* function (equation 4) on the basis of secret key $K^{key}$ which produces four encrypted message blocks ($B_1$, $B_2$, $B_3$, and $B_4$). The $I^{plane}$ of HSI image $I^{HSI}$ is divided into four sub-images and are rotated at different angles using secret key $K^{key}$ via function 5 (*ImageSubDivision*) which results in four sub-images ($I_1$, $I_2$, $I_3$, and $I_4$). Each message block is embedded in its corresponding sub-image using magic LSB method of equation 6. Finally, equation 7 (*ReconstructStego*) re-rotates the sub-stego images to form the stego $I^{plane}$ which is then combined with $H^{plane}$ and $S^{plane}$ to construct the stego image $I^S$. The receiver has to apply the reverse operations in order to extract the original secret information. The major steps of the proposed framework are shown in Figure 3.

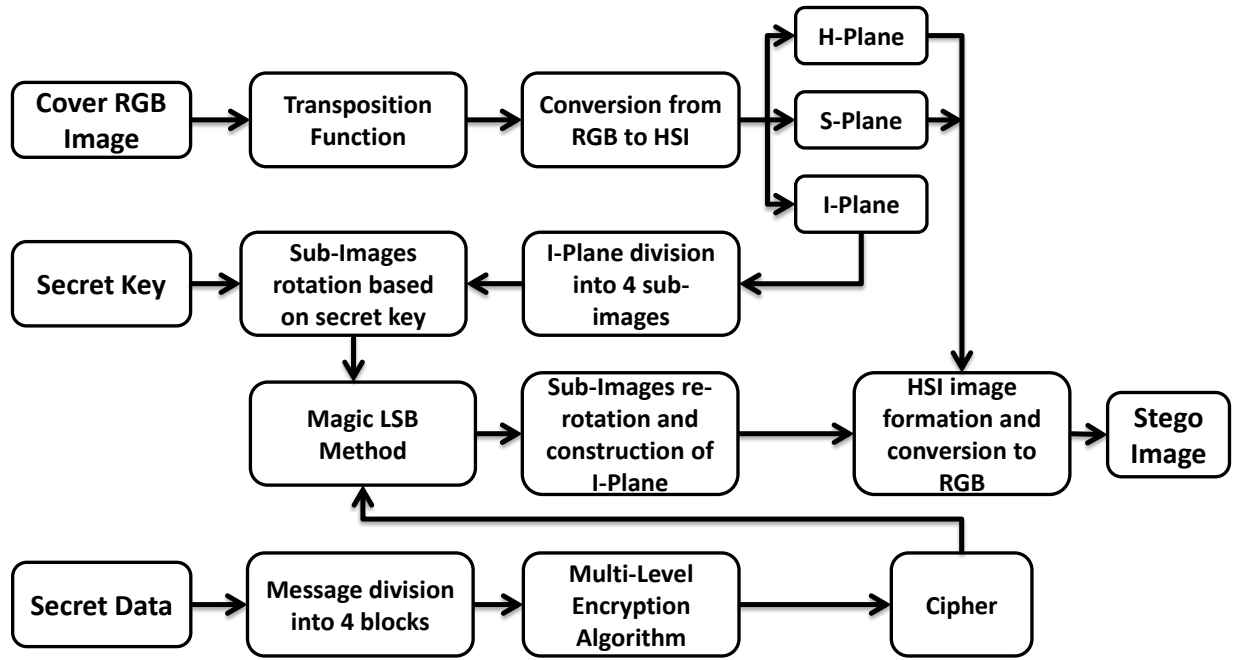

Figure 3: The proposed steganographic model

The MLEA, embedding algorithm, and extraction algorithm are described in detail in the subsequent sections, understanding the conceptual novelty of the proposed scheme. MLEA is an encryption algorithm consisting of different operations that encrypts secret data based on a stego key and produces four distinct encrypted blocks of message. Embedding algorithm embeds the encrypted secret information into the input image and extraction algorithm extracts the hidden data from the stego image.

### 3.1 Multi-Level Encryption Algorithm (MLEA)
The MLEA encrypts the secret data before it is embedded into the carrier image. This algorithm applies different encryption operations on secret data, increasing its security. The main steps of MLEA are given in Algorithm 1:

---
**Algorithm 1.** Multi-Level Encryption Algorithm
---
**Input**: Secret Data (D) and Stego Key ($K^{key}$)
1. **Initialize** K ←key, D ←secret data, Dsize← size(D), $B_1$← Dsize/4, $B_2$← $B_1$, $B_3$← $B_2$, $B_4$← $B_3$, M← (length(D)*8), KB← (length(K)*8) and j=8
2. **for** each character D(i) and K(i) in secret data D and stego key K **do**
   a. Convert D(i) into 8-bits and concatenate it with M.
   b. Convert K(i) into 8-bits and concatenate it with KB.
   **end for**
3. **for** i←1 to size(M), **do**
   a. temp← M(i:i+7);
   b. $B_1$ ← temp(8) & temp(1);
   c. $B_2$← temp(7) & temp(2);

    d.  $B_3 \leftarrow$ temp(6) & temp(3);
    e.  $B_4 \leftarrow$ temp(5) & temp(4);
    f.  $i \leftarrow i+8$;
   **end for**
 4. **for** each and every bit $B_1(i)$, $B_2(i)$, $B_3(i)$, $B_4(i)$, and $KB(i)$, **do**
    a.  $B_1(i) \leftarrow (B_1(i) \oplus$ logical 1);
    b.  $B_2(i) \leftarrow (B_2(i) \oplus$ logical 1);
    c.  $B_3(i) \leftarrow (B_3(i) \oplus$ logical 1);
    d.  $B_4(i) \leftarrow (B_4(i) \oplus$ logical 1);
    e.  $KB(i) \leftarrow (KB(i) \oplus$ logical 1);
   **end for**
 5. **for** each 8-bits combination in $B_1$ **do**
    **for** $i \leftarrow 1$ to 4 **do**
     a. tempVar $\leftarrow B_1(i)$;
     b. $B_1(i) \leftarrow B_1(j)$;
     c. $B_1(j) \leftarrow$ tempVar;
     d. $j \leftarrow j-1$;
    **end for**
    $j \leftarrow 8$;
   **end for**
 6. Repeat Step 5 for $B_2$, $B_3$, $B_4$, and $KB$.
 7. **for** each bit $KB(i)$ in $KB$ **do**
    **if** $KB(i)=1$ **then**
     $B_1(i) \leftarrow (B_1(i) \oplus$ logical 1));
    **else**
     $B_1(i) \leftarrow B_1(i)$;
    **end if-else**
   **end for**
 8. Repeat Step 7 for $B_2$, $B_3$, and $B_4$.

**Output**: Cipher in four message blocks ($B_1$, $B_2$, $B_3$, and $B_4$)

### 3.2 Embedding Algorithm

The embedding algorithm is based on color model conversion from RGB to HSI and the magic LSB method. The cover image is transposed and converted to HSI color space. The I-plane of transposed HSI image is divided into four sub-images and each sub-image is rotated at a certain angle based on secret key. The encrypted message of MLEA in four distinct blocks is then hidden using magic LSB method in the rotated four sub-images. The main steps of the proposed embedding algorithm are given in Algorithm 2:

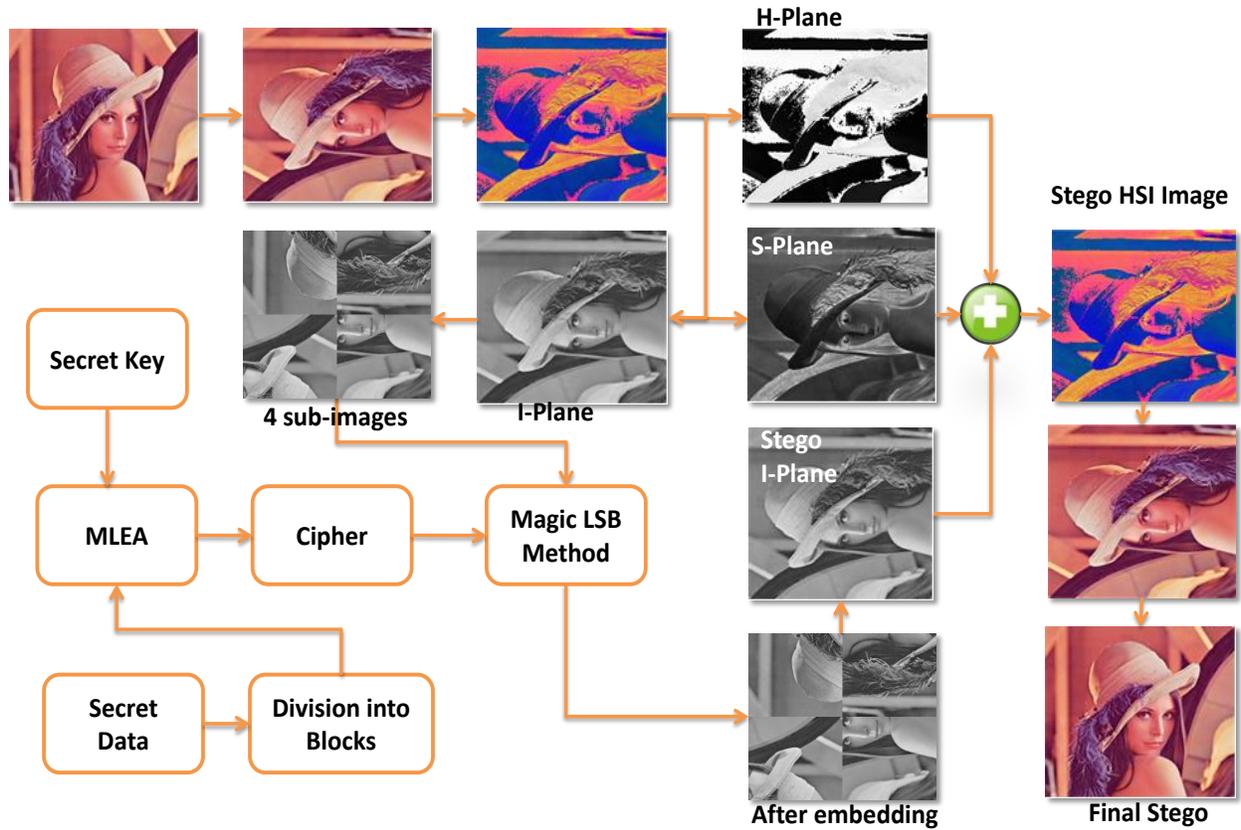

Figure 4: Detailed pictorial representation of the proposed scheme

---

**Algorithm 2.** Embedding Algorithm

**Input**: Cover Color Image ($I^C$), Secret data ($D$), Secret key ($K^{key}$)

1. **Initialize** $I^C \leftarrow$ cover image, $D \leftarrow$ secret data, $K^{key} \leftarrow$ secret key
2. Encrypt $D$ using MLEA (algorithm 1) to get four distinct blocks $B_1$, $B_2$, $B_3$, and $B_4$.
3. Apply the transposition function to transpose $I^C$ and get the transposed image $I^T$.
4. Transform the image from RGB to HSI and separate the achromatic plane (I-plane).
5. Divide the I-plane into 4 sub-images of equal size i.e. $I_{c1}$, $I_{c2}$, $I_{c3}$, and $I_{c4}$.
6. Rotate the sub-images at certain angles using secret key $K^{key}$.
7. Embed each message block to its corresponding image block using magic LSB as:
8. Generate a magic matrix (MGM) of size equal to the size of sub-image.
9. **While** counter <= size of message block **do**
   a. Consider a pixel $I_j (x, y)$ (here $j$ shows the block number)
   b. Find the index of a particular message bit in MGM.
   c. Replace the LSB of the pixel at that particular index in sub-image block
   d. counter $\leftarrow$ counter +1;
   **end**
10. Repeat Step 8 and Step 9 for the remaining 3 sub-image blocks.
11. Re-rotate the sub-images and combine it to form stego I-plane.
12. Combine $H^{plane}$, $S^{plane}$, and $I^{plane}$; convert the HSI image to RGB, and transpose it to

get the final stego image.

**Output**: Stego Image ($I^S$)

---

The magic LSB method is further explained using a simple example, considering a cover image $I^C$= {40, 56, 21, 55, 65, 52, 44, 78 and 79} and secret bits $B^s$= (01000001)$_2$. To embed this sequence of bits, first we generate a magic matrix of size equal to the size of stego image i.e. 3×3. The reasons behind the magic matrix used for message embedding are given in Table 3:

Table 3: Properties of Magic Matrix

| | |
|---|---|
| i. | Magic matrix contains unique numbers (non-repeated) |
| ii. | The numbers inside a given magic matrix are not greater than the product of its rows and columns. ( In the case of 3×3 matrix, every number will be equal or less than 9 as given) |
| iii. | The sum of all rows, columns and its diagonals are equal to the same number ( In the case of 3×3, the sum is 15 i.e. 8+1+6=3+5+7=4+9+2=8+3+4=1+5+9=6+7+2=**15**. Similarly 8+5+2=4+5+6=**15** (diagonals) |

The 3×3 cover image ($I^C$), magic matrix (MGM), and stego image ($I^S$) are:

$$I^C: \begin{bmatrix} 40 & 56 & 21 \\ 55 & 65 & 52 \\ 44 & 78 & 79 \end{bmatrix} \quad MGM: \begin{bmatrix} 8 & 1 & 6 \\ 3 & 5 & 7 \\ 4 & 9 & 2 \end{bmatrix} \quad \text{After hiding: } I^S: \begin{bmatrix} \mathbf{41} & 56 & \mathbf{20} \\ \mathbf{54} & \mathbf{64} & 52 \\ 44 & 78 & 79 \end{bmatrix}$$

The magic matrix shows the location where we have to store the secret bits i.e. the first secret bit will be embedded in 56 (row 1, column 2), 2$^{nd}$ bit in 79 (row 3, column 3), 3$^{rd}$ bit in 55 (row 2, column 1), 4$^{th}$ bit in 44 (row 3, column 1, 5$^{th}$ bit in 65 (row 2, column 2), 6$^{th}$ bit in 21 (row 1, column 3), 7$^{th}$ bit in 52 (row 2, column 3), and 8$^{th}$ bit in 40 (row 1, column 1), and so on. The numbers shown in bold face in $I^S$ are changed as a result of embedding. This process disperses the encrypted secret bits in each sub-image, hence makes its extraction more challenging for attackers.

### 3.3 Extraction Algorithm

The extraction algorithm transposes the stego image and then converts it from RGB to HSI color space. The I-plane of the converted stego image is decomposed into 4 sub-images. Each sub-image is rotated at certain angles as rotated in embedding algorithm based on secret key. The next step is to extract the messages from each sub-image and, then these messages are decrypted to get the actual secret message. The major steps of extraction algorithm are given in Algorithm 3:

---

**Algorithm 3.** Extraction Algorithm

**Input**: Stego Image ($I^S$), Secret key ($K^{key}$)
1. **Initialize** $I^S$ ←stego RGB image, $K^{key}$← secret key
2. Apply the transposition function to transpose the image $I^S$ to $I^T$.
3. Transform the image from RGB to HSI and separate the achromatic plane (I-plane.)

4. Divide the I-plane into 4 sub-images of equal size i.e. $I_{S1}$, $I_{S2}$, $I_{S3}$, and $I_{S4}$.
5. Rotate the sub-images at certain angles using secret key $K^{key}$.
6. Generate a magic matrix (MGM) of size equal to the size of sub-image.
7. **While** size of message block >= counter **do**
    a. Consider a pixel $I_{Sj}(x, y)$ (here j shows the block number)
    b. Extract the LSB of the pixel in sub-image which is located at the first coming index in MGM (Start from index 1 and continue it up to end of message size)
    **end while**
8. Repeat Step 6 and Step 7 for the remaining 3 sub-image blocks to get 4 message blocks.
9. Apply the reverse operations of MLEA on message blocks to get the decrypted bits.
10. Convert the bits into actual data form.

**Output**: Secret data (D)

## 4. Experimental Results and Discussion

In this section, we present the detail of the experimental setup for the proposed method and other existing discussed methods. The proposed technique is compared with seven state of the art techniques whose brief description is given in the next sub-section 4.1. All the mentioned techniques are simulated using MATLAB R2013a. A number of different experiments were conducted based on multiple image quality assessment metrics (IQAMs) [49-51], assessing the effectiveness of the proposed scheme. The following sub-sections present the experimental results and critical discussions in detail.

**4.1 Description of Steganographic Methods with which the Proposed Method is Compared**

The proposed method is compared with seven existing methods including classical LSB substitution method, stego color cycle (SCC)[41], pixel indicator technique (PIT)[18], five modulus method (FMM)[52], Karim's technique[53], our first recently published cyclic steganographic technique (CST)[42] and our $2^{nd}$ simple HSI (SHSI) technique[5]. The LSB method, cyclic LSB method, PIT, and CST are already discussed in the literature review section. The FMM method divides the host image into a set of blocks with block sizes equal to k×k pixels where k shows the window size. Each pixel in the window is then modified such that it becomes divisible by 5. Although the proposed method scatters data in the whole image; its payload is limited i.e. less than 1bpp in many cases. Karim's [53] method embeds secret data in GREEN or BLUE channel of the carrier image's pixels, increasing the security. The decision about which channel to use for embedding is based on LSB of RED channel and secret key bits. The RED channel LSB and secret key bit is xored and then a decision is taken on the basis of its result to replace the LSB of GREEN or BLUE channel. Our SHSI method is based on color model exchange. It transforms the image from RGB to HSI color space and hides data directly via simple LSB method.

**4.2 Dataset**

In this sub-section, the datasets of the images and the sources from where they were taken have been presented. Two datasets referred to as the USC-SIPI-ID[54] and COREL Database [55] consisting of standard color images were used for assessing the performance of mentioned schemes and the proposed scheme. Fifty images including different edgy and smooth color images of dimension 512×512 were taken from USC-SIPI-ID dataset, consisting of Lena, mandrill (baboon), peppers, trees, and house etc. In the same context, one hundred color images were selected for evaluation from the COREL database with dimension 384×256 pixels. These images are adjusted to dimension (256×256) for consistency. In this paper, a total of one hundred and fifty (150) images have been used for analysis of existing mentioned and the proposed techniques.

### 4.3 Quantitative Evaluation
This sub-section demonstrates the complete procedure of quantitative analysis that has been used in this paper. All the mentioned techniques are experimentally evaluated from three different perspectives based on multiple IQAMs whose detail is given in Table 4.

Table 4: Types of experiments for evaluation of the proposed algorithm

| Experiment # | Cipher Size | Size in Pixels | Images |
|---|---|---|---|
| *Perspective 1* | Equal (8KB) | 256×256 | Different |
| *Perspective 2* | Variable (2KB, 4KB, 6KB & 8KB | 256×256 | Same |
| *Perspective 3* | Equal (8KB) | Variable (128×128), (256×256), (512×512), (1024×1024) | Same |

According to *perspective* 1, a text file of 8KB is embedded in different edgy and smooth color images having size 256×256 in pixels. A total of 150 images were tested using *perspective* 1. The second *perspective* is to encode text files of different sizes in the same images of uniform dimension (256×256 in pixels). In third *perspective*, multiple color images with different resolutions (128×128, 256×256, 512×512 and 1024×1024) were used. The size of the cipher in this experiment is the same as *perspective* 1 i.e. 8KB. The detailed experimental results of these three perspectives are shown in section 4.3.1.

### 4.3.1 Quantitative Results and Discussion
This sub-section presents the comparison of the proposed approach and the other seven existing schemes: classical LSB method, SCC[41], PIT[18], FMM[52], Karim's approach[53] and our two recently published approaches including CST[42], and SHSI[5]. The comparison is based on well-known IQAMs[56] including peak signal-to-noise ratio (PSNR), normalized cross correlation (NCC)[57], structural similarity index (SSIM)[58], and mean absolute error (MAE). These metrics are computed using equation 8-12 respectively as follows:

$$PSNR = 10 \log_{10} \left( \frac{C_{\max}^2}{MSE} \right) \tag{8}$$

$$MSE = \frac{1}{MN} \sum_{x=1}^{M} \sum_{y=1}^{N} (S_{xy} - C_{xy})^2 \tag{9}$$

$$SSIM(C,S) = \frac{(2\mu_x\mu_y + C_1)(2\sigma_{xy} + C_2)}{(\mu_x^2 + \mu_y^2 + C_1)(\sigma_x^2 + \sigma_y^2 + C_2)} \tag{10}$$

$$NCC = \frac{\sum_{x=1}^{M}\sum_{y=1}^{N}(S(x,y) \times C(x,y))}{\sum_{x=1}^{M}\sum_{y=1}^{N}(S(x,y))^2} \quad (11)$$

$$MAE = \left(\frac{1}{N}\right)\sum_{x=1}^{N}|C_x - S_x| \quad (12)$$

Note that *M* and *N* show image dimensions, *x* and *y* are loop counters, *C* is cover image, *S* is stego image, and $C_{max}$ is the maximum pixel intensity among both images. $\sigma_x$, $\sigma_y$, $\sigma_{xy}$, $\mu_y$, and $\mu_x$ refer to some local parameters that are related to statistics[59, 60].

A few sample images from the datasets for quantitative experiments are shown in Figures 5-8. The incurred results of all mentioned algorithms based on PSNR, SSIM, NCC, and MAE from three different perspectives are listed in Table 5-11 respectively.

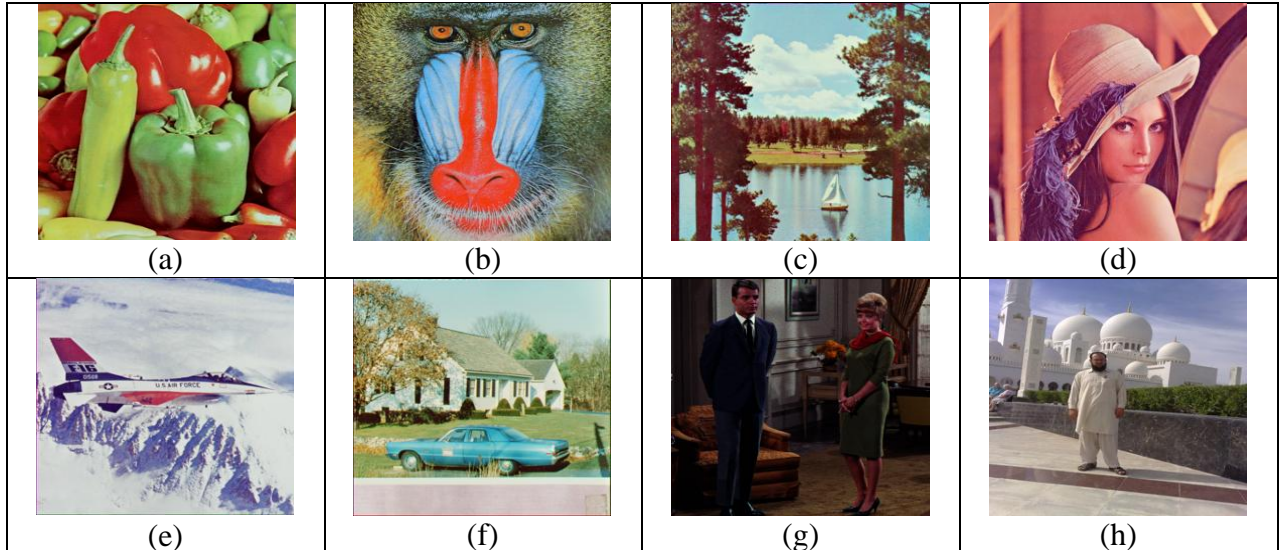

Figure 5: *Perspective* 1; Sample test cover images from the dataset; (a) Peppers (b) Baboon (c) Trees (d) Lena (e) F16jet (f) House (g) Couple (h) Scene.

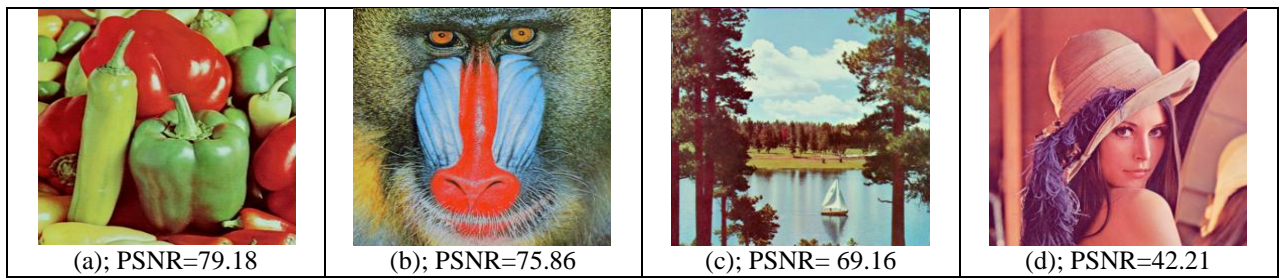

(a); PSNR=79.18    (b); PSNR=75.86    (c); PSNR= 69.16    (d); PSNR=42.21

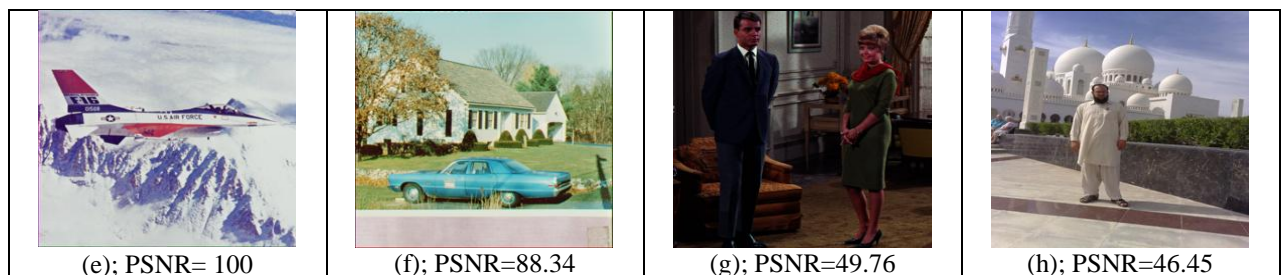

(e); PSNR= 100    (f); PSNR=88.34    (g); PSNR=49.76    (h); PSNR=46.45

Figure 6: *Perspective* 1; A few sample test stego images from the dataset produced by the proposed method, each containing 8KB cipher; (a) Peppers (b) Baboon (c) Trees (d) Lena (e) F16jet (f) House (g) Couple (h) Scene.

Table 5: *Perspective* 1 Results; Comparison of the proposed method with existing seven methods based on PSNR (dB) by hiding same amount of cipher (8KB) in different images of same resolution (256×256)

| Serial # | Image Name | Classic LSB Method | SCC[41] Method | PIT [18] | FMM [52] | CST [42] | SHSI [5] | Karim's Method[53] | Proposed Method |
|---|---|---|---|---|---|---|---|---|---|
| 1 | Peppers | 55.83 | 49.82 | 48.52 | 45.77 | 16.07 | 78.45 | 49.72 | 79.18 |
| 2 | Baboon | 54.73 | 47.97 | 46.89 | 44.55 | 48.95 | 75.70 | 47.90 | 75.86 |
| 3 | House | 52.04 | 52.89 | 51.07 | 67.55 | 51.17 | 83.57 | 52.79 | 88.34 |
| 4 | Trees | 56.27 | 49.76 | 48.60 | 46.12 | 38.54 | 69.30 | 49.73 | 69.16 |
| 5 | Lena | 42.51 | 42.60 | 42.30 | 43.57 | 55.92 | 42.18 | 42.56 | 42.21 |
| 6 | Moon | 56.02 | 47.26 | 46.39 | 45.82 | 47.49 | 78.54 | 47.25 | 77.62 |
| 7 | Scene | 46.11 | 45.06 | 44.01 | 42.88 | 28.53 | 46.63 | 45.08 | 46.45 |
| 8 | Couple | 48.40 | 47.91 | 46.58 | 46.25 | 55.91 | 51.04 | 47.92 | 49.76 |
| 9 | Design1 | 45.97 | 46.14 | 45.42 | 41.24 | 46.41 | 46.48 | 46.40 | 46.40 |
| 10 | Competition1 | 45.23 | 42.41 | 41.45 | 40 | 34.04 | 43.55 | 42.28 | 43.66 |
| 11 | Baboon3 | 41.22 | 39.05 | 38.58 | 39.16 | 22.02 | 42.25 | 39.10 | 42.10 |
| 12 | F16jet | 52.35 | 53.41 | 51.29 | 76.72 | 47.48 | 100 | 49.80 | 100 |
| 13 | Building1 | 43.34 | 43.45 | 43.13 | 64.18 | 28.84 | 100 | 43.44 | 100 |
| 14 | Corel_141 | 44.63 | 40.24 | 40.09 | 40.03 | 40.24 | 48.08 | 40.14 | 48.36 |
| 15 | Corel_134 | 43.24 | 40.35 | 39.72 | 39.12 | 40.35 | 44.19 | 40.48 | 44.46 |
| 16 | Corel_205 | 41.42 | 39.14 | 38.72 | 39.05 | 39.14 | 42.06 | 39.10 | 42.15 |
| 17 | Corel_130 | 45.36 | 43.45 | 42.76 | 41.94 | 43.45 | 54.40 | 43.41 | 54.87 |
| 18 | Corel_118 | 44.75 | 42.85 | 41.38 | 40.7 | 42.85 | 45.23 | 42.55 | 45.66 |
| 19 | Corel_301 | 41.73 | 36.40 | 36.38 | 41.32 | 36.40 | 45.16 | 36.41 | 45.32 |
| 20 | Corel_392 | 37.89 | 36.46 | 36.34 | 39.09 | 36.46 | 42.30 | 36.41 | 42.20 |
| 21 | Corel_300 | 37.64 | 36.70 | 36.52 | 39.71 | 36.70 | 41.91 | 36.65 | 41.76 |
| 22 | Corel_143 | 49.11 | 44.90 | 43.89 | 44.52 | 44.90 | 54.93 | 44.74 | 55.09 |
| 23 | Corel_138 | 46.71 | 44.35 | 43.71 | 42.62 | 44.35 | 52.41 | 44.38 | 52.34 |
| 24 | Corel_388 | 38.08 | 35.96 | 35.85 | 39.41 | 35.96 | 42.70 | 35.94 | 42.71 |
| 25 | Corel_397 | 38.10 | 34.65 | 34.51 | 37.40 | 34.65 | 40.18 | 34.63 | 40.16 |
| **Avg. of 150 images** | | **45.28** | **41.83** | **41.22** | **41.97** | **37.38** | **47.97** | **41.78** | **47.93** |

Table 6: *Perspective* 1 Results; SSIM based comparison of the proposed scheme with existing seven schemes

| Serial# | Image name | Classic LSB Method | SCC[41] Method | PIT[18] | FMM[52] | CST[42] | SHSI [5] | Karim's Method[53] | Proposed Method |
|---|---|---|---|---|---|---|---|---|---|
| 1 | Lena | 0.9981 | 0.9989 | 0.9971 | 0.9822 | 0.9993 | 0.9994 | 0.9989 | 0.9994 |
| 2 | Baboon | 0.9989 | 0.9993 | 0.9985 | 0.9925 | 0.995 | 0.9998 | 0.9992 | 0.9998 |
| 3 | Couple | 0.9967 | 0.9985 | 0.9936 | 0.9775 | 0.997 | 0.9992 | 0.998 | 0.9992 |
| 4 | Trees | 0.9964 | 0.997 | 0.9956 | 0.9858 | 0.998 | 0.9995 | 0.997 | 0.9995 |
| 5 | Baboon2 | 0.9953 | 0.9938 | 0.9928 | 0.9888 | 0.874 | 0.9998 | 0.9937 | 0.9998 |
| 6 | Peppers | 0.8843 | 0.8774 | 0.8756 | 0.9488 | 0.989 | 0.9994 | 0.8773 | 0.9994 |
| 7 | Scene | 0.9979 | 0.9989 | 0.997 | 0.9817 | 0.9909 | 0.9996 | 0.9988 | 0.9996 |
| 8 | House | 0.9983 | 0.999 | 0.9974 | 0.986 | 0.9904 | 0.9995 | 0.9989 | 0.9995 |
| 9 | Scene3 | 0.9989 | 0.9994 | 0.9983 | 0.9895 | 0.6690 | 0.9997 | 0.9993 | 0.9997 |
| 10 | Design2 | 0.6885 | 0.6699 | 0.6677 | 0.9916 | 0.9504 | 0.9991 | 0.6699 | 0.9991 |
| **Average of 150 images** | | **0.9689** | **0.9560** | **0.9543** | **0.9751** | **0.9560** | **0.9989** | **0.9582** | **0.9995** |

Table 7: *Perspective* 1 Results; Comparison of the proposed scheme with existing seven schemes based on NCC

| Serial# | Image name | Classic LSB Method | SCC[41] Method | PIT[18] | FMM[52] | CST[42] | SHSI [5] | Karim's Method[53] | Proposed Method |
|---|---|---|---|---|---|---|---|---|---|
| 1 | F16jet | 0.9997 | 0.9997 | 0.9996 | 0.9993 | 0.9993 | 0.9994 | 0.9997 | 0.9993 |
| 2 | Building1 | 0.9795 | 0.9795 | 0.9795 | 0.9993 | 0.995 | 0.9998 | 0.9796 | 0.9994 |
| 3 | Baboon | 0.9998 | 0.9998 | 0.9997 | 0.999 | 0.997 | 0.9992 | 0.9998 | 0.9995 |
| 4 | House | 0.9999 | 0.9999 | 0.9998 | 0.9994 | 0.998 | 0.9995 | 0.9999 | 0.9996 |
| 5 | Trees | 0.999 | 0.999 | 0.9989 | 0.9997 | 0.874 | 0.9998 | 0.999 | 0.9994 |
| 6 | Moon | 0.9998 | 0.9998 | 0.9997 | 0.999 | 0.989 | 0.9994 | 0.9998 | 0.9994 |
| 7 | Lena | 0.9999 | 1 | 0.9999 | 0.9994 | 0.9909 | 0.9996 | 1 | 0.9993 |
| 8 | Parrot | 0.9999 | 0.9991 | 0.999 | 0.9985 | 0.9904 | 0.9995 | 0.9991 | 0.9997 |
| 9 | Laserlight | 0.9967 | 0.9938 | 0.9937 | 0.9992 | 0.6690 | 0.9997 | 0.9938 | 0.9993 |
| 10 | Kite | 0.9762 | 0.9582 | 0.9582 | 0.9974 | 0.9504 | 0.9991 | 0.9582 | 0.9996 |
| **Average of 150 images** | | **0.9668** | **0.9529** | **0.9529** | **0.9984** | **0.9560** | **0.9989** | **0.9559** | **0.9989** |

Table 8: *Perspective* 1 Results; MAE based comparison of the proposed scheme with mentioned seven schemes

| Serial# | Image name | Classic LSB Method | SCC[41] Method | PIT[18] | FMM[52] | CST[42] | SHSI [5] | Karim's Method[53] | Proposed Method |
|---|---|---|---|---|---|---|---|---|---|
| 1 | Lena2 | 0.0772 | 0.0772 | 0.1908 | 0.9964 | 0.0672 | 0.0738 | 0.0768 | 0.0008 |
| 2 | Parrot | 0.0774 | 0.0766 | 0.1883 | 0.9901 | 0.0746 | 0.0737 | 0.0757 | 0.0011 |
| 3 | Laserlight | 0.0766 | 0.0764 | 0.1914 | 1.0009 | 0.0764 | 0.0746 | 0.0766 | 0.0008 |
| 4 | Kite | 0.0761 | 0.0763 | 0.1851 | 0.9851 | 0.0743 | 0.0738 | 0.0748 | 0.0002 |
| 5 | Rose | 0.0772 | 0.0769 | 0.1904 | 1.0024 | 0.0779 | 0.0792 | 0.0762 | 0.0013 |
| 6 | Competition | 0.0671 | 0.0663 | 0.1935 | 0.8546 | 0.0653 | 0.0639 | 0.0669 | 0.0289 |
| 7 | Scene | 0.077 | 0.0773 | 0.1906 | 1.0001 | 0.0783 | 0.0737 | 0.0767 | 0.0001 |
| 8 | Hackers | 0.0726 | 0.0731 | 0.1879 | 0.9235 | 0.0741 | 0.043 | 0.073 | 0.0127 |
| 9 | Scene3 | 0.0762 | 0.0770 | 0.1898 | 1.0018 | 0.047 | 0.0468 | 0.0768 | 0.0011 |
| 10 | Design2 | 0.0200 | 0.0680 | 0.1278 | 0.6536 | 0.0638 | 0.0659 | 0.0669 | 0.0090 |
| **Average of 150 images** | | **0.0740** | **0.0756** | **0.1843** | **0.9645** | **0.0750** | **0.0746** | **0.0752** | **0.0043** |

Table 5-8 shows the experimental results of the proposed scheme and the other seven schemes based on various IQAMs using *perspective* 1. According to *perspective* 1, equal size of text

(8KB) is encoded in different diverse images of the same resolution (256×256). The anticipated scheme clearly dominates the existing seven schemes by attaining highest values of the mentioned IQAMs. The last line of Table 5-8 shows the average value of each metric computed over one hundred and fifty images (150). The average results demonstrated in the last row of Table 5-8 in bold face clearly show the excellence of the proposed scheme as compared to the other seven mentioned approaches.

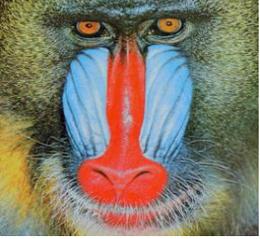
Cipher=2KB
PSNR=84.39

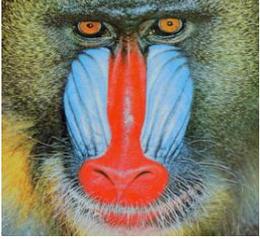
Cipher=4KB
PSNR=77.58

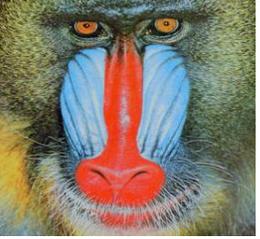
Cipher=6KB
PSNR=75.57

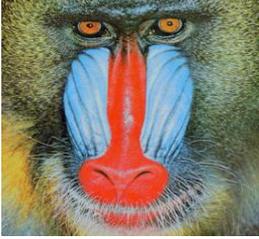
Cipher=8KB
PSNR=75.86

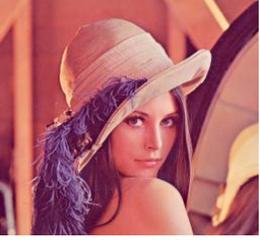
Cipher=2KB
PSNR=56.69

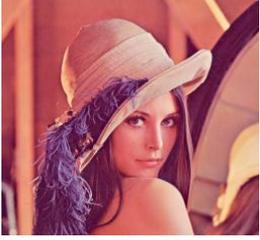
Cipher=4KB
PSNR=54.62

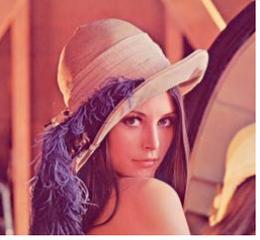
Cipher=6KB
PSNR=53.29

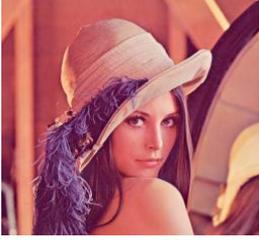
Cipher=8KB
PSNR=52.42

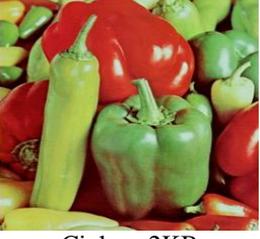
Cipher=2KB
PSNR=86.54

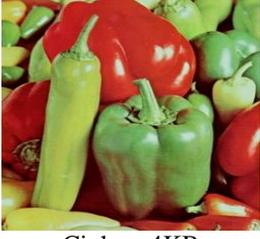
Cipher=4KB
PSNR=82.43

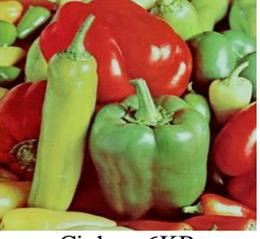
Cipher=6KB
PSNR=79.36

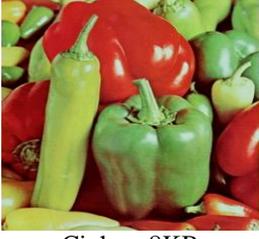
Cipher=8KB
PSNR=79.18

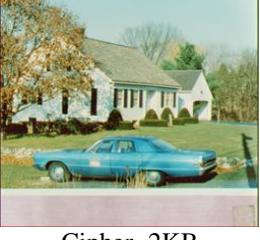
Cipher=2KB
PSNR=86.12

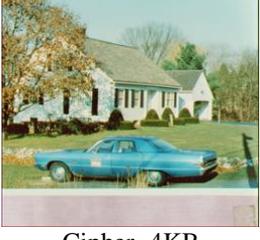
Cipher=4KB
PSNR=84.36

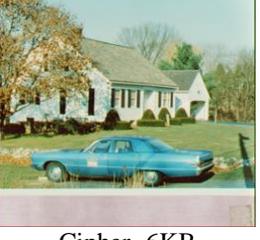
Cipher=6KB
PSNR=83.57

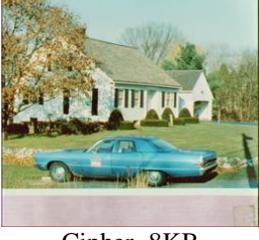
Cipher=8KB
PSNR=88.34

Figure 7: Sample stego images from dataset for *perspective* 2. First row shows baboon image with 2KB, 4KB, 6KB, and 8KB cipher. Second row shows Lena image with different amount of data. Third row presents pepper images and fourth row depicts different versions of house image.

Table 9: *Perspective* 2 results; Comparison of the proposed scheme with other seven mentioned algorithms based on PSNR (dB) with variable amount of cipher embedded in same images of same dimensions (256×256)

| Image Name | Secret data | Cipher size in | Classic LSB | SCC Method | PIT | FMM | CST | SHSI | Karim's Method | Proposed Method |
|---|---|---|---|---|---|---|---|---|---|---|

| | (KBs) | bytes | | | | | | | | |
|---|---|---|---|---|---|---|---|---|---|---|
| Baboon image with dimension 256×256 | 2 | 2406 | 60.46 | 48.40 | 48.58 | 44.57 | 49.64 | 80.29 | 48.39 | 84.39 |
| | 4 | 4177 | 57.42 | 48.27 | 47.80 | 44.58 | 49.38 | 79.12 | 48.21 | 77.58 |
| | 6 | 6499 | 55.68 | 48.10 | 46.98 | 44.57 | 49.13 | 77.91 | 48.03 | 75.57 |
| | 8 | 8192 | 54.73 | 47.97 | 46.89 | 44.57 | 48.95 | 75.70 | 47.90 | 75.86 |
| | **Average** | | **57.07** | **48.18** | **47.56** | **44.57** | **49.27** | **78.25** | **48.13** | **78.35** |
| Lena with resolution 256×256 | 2 | 2406 | 46.23 | 46.29 | 44.32 | 46.12 | 61.77 | 42.42 | 46.27 | 56.69 |
| | 4 | 4177 | 49.58 | 49.89 | 44.07 | 46.13 | 58.75 | 42.30 | 49.84 | 54.62 |
| | 6 | 6499 | 49.32 | 49.75 | 43.92 | 46.13 | 56.95 | 42.42 | 49.68 | 53.29 |
| | 8 | 8192 | 49.14 | 49.65 | 42.30 | 46.13 | 55.92 | 42.18 | 49.57 | 52.42 |
| | **Average** | | **48.57** | **48.90** | **43.65** | **46.13** | **58.35** | **42.33** | **48.84** | **54.25** |
| Peppers image with dimension 256×256 | 2 | 2406 | 61.59 | 50.13 | 50.93 | 45.77 | 50.05 | 87.53 | 50.11 | 86.54 |
| | 4 | 4177 | 58.66 | 50.03 | 50.10 | 45.76 | 49.93 | 82.52 | 49.95 | 82.43 |
| | 6 | 6499 | 56.84 | 49.91 | 49.42 | 45.76 | 49.79 | 82.52 | 49.83 | 79.36 |
| | 8 | 8192 | 55.83 | 49.82 | 48.52 | 45.76 | 49.70 | 80.26 | 49.72 | 79.18 |
| | **Average** | | **58.23** | **49.97** | **49.74** | **45.77** | **49.86** | **83.21** | **49.90** | **81.88** |
| House image with resolution 256×256 | 2 | 2406 | 53.43 | 53.74 | 53.32 | 67.49 | 53.74 | 100 | 53.71 | 86.12 |
| | 4 | 4177 | 47.79 | 53.39 | 53.84 | 67.53 | 53.39 | 89.13 | 53.33 | 84.36 |
| | 6 | 6499 | 52.37 | 53.09 | 53.01 | 67.39 | 53.09 | 85.70 | 53.02 | 83.57 |
| | 8 | 8192 | 52.04 | 52.89 | 51.07 | 67.34 | 52.89 | 83.57 | 52.79 | 88.34 |
| | **Average** | | **51.41** | **53.28** | **52.81** | **67.44** | **51.17** | **89.60** | **53.21** | **85.59** |

Table 10: *Perspective* 2 results; NCC based comparison of the proposed scheme with other seven mentioned algorithms

| Image Name | Secret data (KBs) | Cipher size in bytes | Classic LSB | SCC Method | PIT | FMM | CST | SHSI | Karim's Method | Proposed Method |
|---|---|---|---|---|---|---|---|---|---|---|
| Lena image with dimension 256×256 | 2 | 2406 | 0.9936 | 0.9996 | 0.9999 | 0.9994 | 0.9996 | 0.9999 | 0.9996 | 1 |
| | 4 | 4177 | 0.9966 | 0.9995 | 0.9996 | 0.9992 | 0.9994 | 0.9998 | 0.9994 | 1 |
| | 6 | 6499 | 0.9946 | 0.9993 | 0.9995 | 0.9990 | 0.9992 | 0.9996 | 0.9995 | 0.9999 |
| | 8 | 8192 | 0.9986 | 0.9991 | 0.9993 | 0.9984 | 0.9990 | 0.9995 | 0.9992 | 0.9999 |
| | **Average** | | **0.9958** | **0.9993** | **0.9995** | **0.999** | **0.9993** | **0.9997** | **0.9994** | **0.9999** |
| Building with resolution 256×256 | 2 | 2406 | 0.9796 | 0.9796 | 0.9795 | 0.9993 | 0.9796 | 0.9999 | 0.9796 | 1 |
| | 4 | 4177 | 0.9794 | 0.9795 | 0.9794 | 0.9993 | 0.9795 | 0.9996 | 0.9795 | 1 |
| | 6 | 6499 | 0.9792 | 0.9793 | 0.9793 | 0.9991 | 0.9793 | 0.9995 | 0.9793 | 1 |
| | 8 | 8192 | 0.9791 | 0.9791 | 0.9791 | 0.9990 | 0.9791 | 0.9993 | 0.9792 | 0.9999 |
| | **Average** | | **0.9793** | **0.9793** | **0.9793** | **0.9991** | **0.9793** | **0.9995** | **0.9794** | **0.9999** |

The experimental results of the mentioned seven algorithms including the proposed approach using *perspective* 2 are listed in Table 9-10. In this type of experiment, some well-known standard color images of dimension (256×256) are selected and different sizes of text is embedded inside it using all the specified methods. These images are chosen for this type of analysis because every new algorithm has to be evaluated by images of different natures (edgy and smooth). For example, the selected images contain the smooth image (Lena), an edgy image (Baboon) and some other images (Peppers, House, and Building, etc.) having a large number of gray levels as compared to the Lena and Baboon images. The average values of PSNR and NCC shown in bold face in Table 9 and Table 10 are much more than the existing mentioned approaches. This distinction illustrates that the proposed approach out-performs in terms of PSNR and NCC as compared to the other mentioned data hiding approaches.

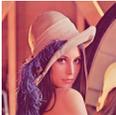
PSNR=58.33
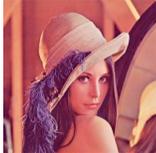
PSNR=52.41
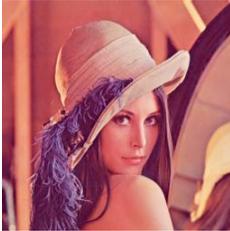
PSNR=57.00
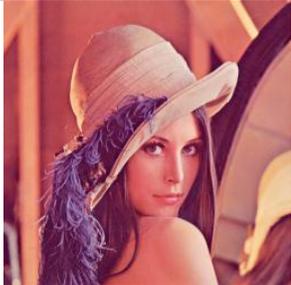
PSNR=59.75

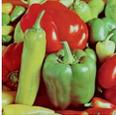
PSNR=100
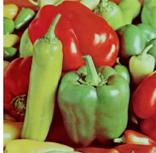
PSNR=79.24
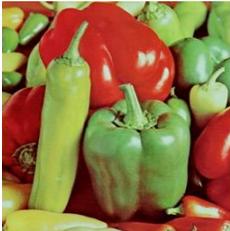
PSNR=87.19
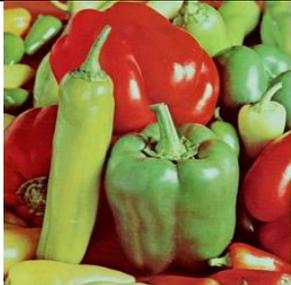
PSNR=90.34

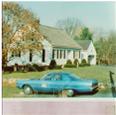
PSNR=69.30
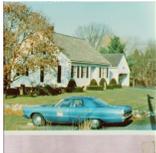
PSNR=64.85
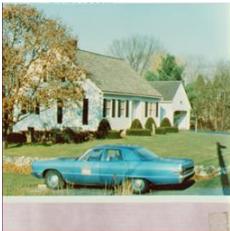
PSNR=63.34
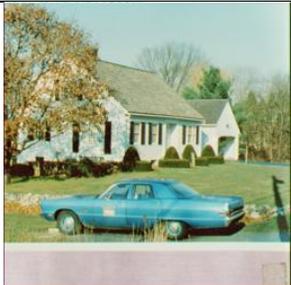
PSNR=72.47

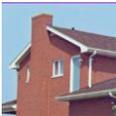
PSNR=63.67
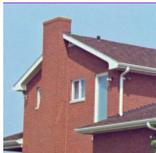
PSNR=62.40
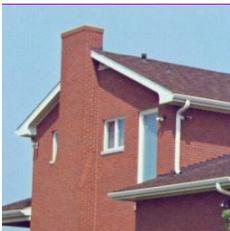
PSNR=59.30
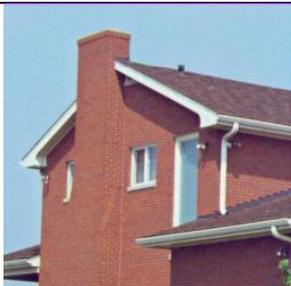
PSNR=65.32

Figure 8: Images dataset for *perspective* 3 containing stego images of different dimensions with their corresponding PSNR scores. Row1 shows Lena images of different resolutions; Row2 is about different versions of pepper image; Row3 depicts the house image with different dimensions; Row4 represents building image with its four versions.

Table 11: *Perspective* 3 results; comparison of the proposed method with other seven methods based on PSNR (dB) by hiding same size of cipher in selected standard images of different resolutions

| Image Name | Image dimensions (in pixels) | Classic LSB Method | SCC Method [41] | PIT[18] | FMM[52] | CST [42] | SHSI [5] | Karim's Method[53] | Proposed Method |
|---|---|---|---|---|---|---|---|---|---|
| Lena image | 128×128 | 42.49 | 42.50 | 45.33 | 45.97 | 42.12 | 58.59 | 42.50 | 58.33 |
| | 256×256 | 49.11 | 49.63 | 50.11 | 46.01 | 47.48 | 52.77 | 49.55 | 52.41 |
| | 512×512 | 49.82 | 49.97 | 50.09 | 46.04 | 48.74 | 57.26 | 49.95 | 57.00 |
| | 1024×1024 | 50.02 | 50.07 | 50.10 | 45.99 | 49.85 | 59.86 | 50.06 | 59.75 |
| | **Average** | **47.86** | **48.04** | **48.90** | **46.00** | **47.05** | **57.12** | **48.01** | **56.87** |
| Peppers image | 128×128 | 64.99 | 50.29 | 48.63 | 45.69 | 50.08 | 87.05 | 50.27 | 100 |
| | 256×256 | 55.88 | 49.74 | 50.23 | 45.77 | 49.59 | 79.34 | 49.68 | 79.24 |
| | 512×512 | 61.88 | 50.06 | 50.19 | 45.76 | 50.01 | 85.77 | 50.05 | 87.19 |
| | 1024×1024 | 67.83 | 50.17 | 50.20 | 45.77 | 50.15 | 100 | 50.16 | 90.34 |
| | **Average** | **62.64** | **50.06** | **49.81** | **45.75** | **49.96** | **88.04** | **50.04** | **89.19** |
| House image | 128×128 | 62.72 | 62.80 | 67.51 | 58.84 | 64.89 | 71.03 | 62.71 | 69.30 |
| | 256×256 | 56.66 | 53.50 | 54.77 | 46.48 | 41.03 | 65.28 | 53.36 | 64.85 |
| | 512×512 | 62.74 | 54.39 | 54.75 | 46.51 | 42.18 | 65.08 | 54.36 | 63.34 |
| | 1024×1024 | 68.82 | 54.69 | 54.79 | 46.54 | 43.14 | 79.61 | 54.68 | 72.47 |
| | **Average** | **62.74** | **56.34** | **57.95** | **49.59** | **47.81** | **70.25** | **56.28** | **67.49** |
| Building image | 128×128 | 76.84 | 78.92 | 55.36 | 61.87 | 64.72 | 64.23 | 77.51 | 63.67 |
| | 256×256 | 49.80 | 50.32 | 47.94 | 48.58 | 47.48 | 62.70 | 47.53 | 62.40 |
| | 512×512 | 50.73 | 50.86 | 51.02 | 46.56 | 47.98 | 59.85 | 50.85 | 59.30 |
| | 1024×1024 | 50.95 | 50.98 | 51.02 | 46.56 | 48.90 | 66.06 | 50.97 | 65.32 |
| | **Average** | **57.08** | **57.77** | **51.34** | **50.89** | **52.27** | **63.21** | **56.72** | **62.67** |

Table 11 illustrates the experimental results of all mentioned approaches using *perspective* 3. In this type of experiment, a text file of 8KB is embedded in four selected color images of different resolutions (128×128, 256×256, 512×512 and 1024×1024 pixels). The incurred results are tabulated in Table 11. By analyzing these results, it can be confirmed that the proposed scheme provides promising results in terms of PSNR in contrast to other mentioned schemes.

### 4.4 Qualitative Analysis

This sub-section briefly illustrates a qualitative analysis that has been used in this paper. HVS has been used for evaluation of the visual quality of stego images of all the presented schemes. A sample of the cover and stego images taken from the Corel database are shown in Figure 9. All these images contain 8KB text that is embedded in the same image of resolution 256×256 using the proposed and the seven other existing schemes except for the image in the first row with label (a). Using naked eye analysis of the stego images, it can be confirmed that there is noticeable distortion in the stego images generated by the existing methods except for the SHSI and the proposed method. The distortion can be noted by comparing the right center portions of the cover and stego images in Figure 9. On the other hand, the stego image with label (j) generated by our proposed algorithm is almost the same to the given cover image with label (a) and there is no obvious distortion between these two images. This means that the stego images generated by our proposed method are of high quality and so it is not easily detectable by the HVS as compared to other methods.

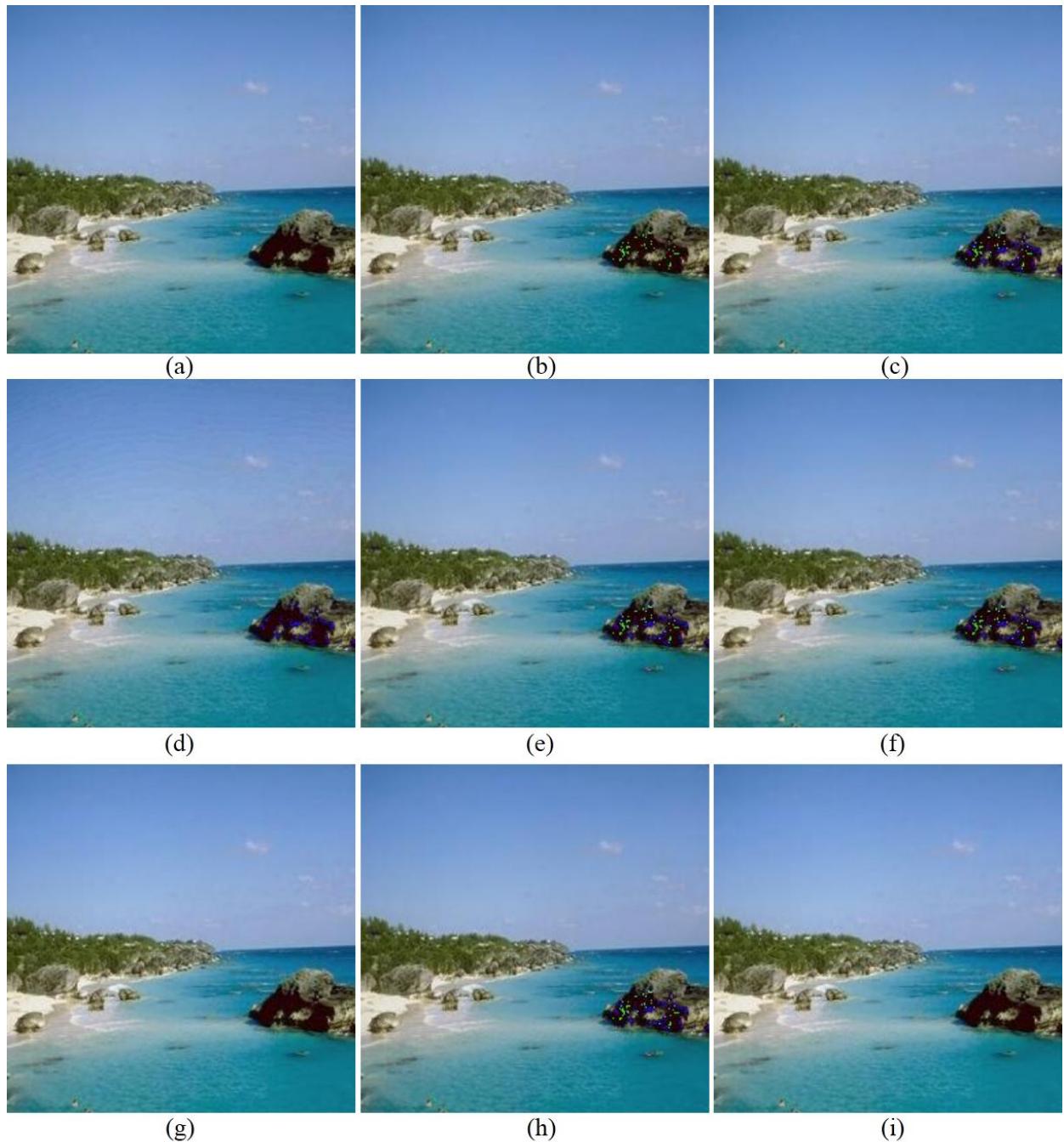

Figure 9: Qualitative analysis using human visual system. Each stego image with dimension (256×256 pixels) contains 8KB cipher except image with label (a). (a) Corel_138 cover image, (b) stego image of classic LSB method with PSNR=46.71, (c) SCC scheme's stego image with PSNR=44.35, (d) stego image of FFM method with PSNR=42.62, (e) Karim's method with stego image PSNR=44.38, (f) CST with stego image of PSNR=44.35, (g) stego image of SHSI method with PSNR=52.41, (h) PIT's stego image PSNR=43.71, and (i) stego image of the proposed method with PSNR=52.34.

**4.5 Performance Analysis**
In this section, the performance of the proposed method and the other competing methods is analyzed and discussed. The performance of a given steganographic algorithm is measured in terms of three well known metrics (capacity/payload, imperceptibility and security). The payload (amount of data to be embedded in the cover image) is the same (1 bits per pixel (bpp)) for all the discussed methods including the proposed method except FMM and PIT. The payload of PIT is greater than all other methods mentioned; however it is less imperceptible and results in the stego images of low quality. The payload of FMM is dependent on the size of a particular window, which is less than 1bpp in many cases, although it disperses the data in different portions of the cover image in the form of small windows.

The classical LSB method is the simplest method and it is easy to hack. SCC hides data in RED, GREEN, and BLUE channels in cyclic form to scatter the data in three channels but it is also easy to crack. CST method uses the concept of randomization to increase the security (how difficult it is for an attacker to extract the hidden data) of the SCC but still extracting data from a few pixels can compromise this method. SHSI transforms the RGB image to HSI and hides the secret data in I-plane using LSB method. SHSI is better than LSB, CST, and SCC in security as it can easily deceive the attacker. On the other hand, SHSI is a highly imperceptible method as compared to the given six methods including the proposed method because it results in stego images of high quality (Table 5, Table 9 and Table 11). Karim's method is more secure as compared to LSB, SCC, CST, PIT, SHSI and FMM because it embeds the secret data in GREEN or BLUE channel by making decision on the XOR result of secret key bits and RED channel LSBs. However its generated stego images are of low quality as compared to CLSB, SCC, PIT and SHSI.

The proposed scheme is better than the existing mentioned schemes in terms of imperceptibility, visual quality and security. The proposed method divides the message into four blocks and encrypts it using MLEA. The image is converted from RGB to HSI; I-plane is divided into four sub-images; each sub-image is rotated at a certain angle using a secret key and finally the distinct four encrypted blocks of message are hidden in four sub-images of I-plane using magic LSB method. These operations make it extremely difficult for attacker to extract the actual hidden data and hence increase the security of the proposed method. In addition to this, the proposed scheme results in high quality stego images and hence it is difficult to detect it using HVS as compared to the other competing methods except SHSI method.

**5. Conclusion and Future Directions**
In this paper, we proposed a novel image steganographic technique (M-LSB-SM) for color images with better imperceptibility and security. The achromatic component of the HSI color model is used instead of an RGB color model, reducing the processing time and increasing the security of hidden data. An average PSNR of 47.93 dB computed over one hundred and fifty images is achieved with this novel approach, which confirms the superiority of the proposed scheme as compared to some other mentioned benchmark schemes. The secret information is divided into four sub-blocks and is passed through MLEA, which makes the attack on this algorithm awful and thus misguides the process of steganalysis. We conclude that our proposed scheme is capable of generating stego images of a sufficient quality that fulfills the favorable demands of modern security systems and users. Our algorithm is simple, easy to implement and a good combination of imperceptibility and security and thus is more feasible to be adopted by steganographic applications.

Although our proposed scheme already demonstrates better results, still some additional improvements are attainable. In future work, we will focus on the following points:
  i. Improving the efficiency of the proposed scheme in terms of payload. Extending MLEA in order to make this approach more powerful.
  ii. Implementing this algorithm in the transform domain to make it resilient against image processing and statistical attacks.

**6. Acknowledgment**
This research is supported by the ICT R&D program of MSIP/IITP. [2014(R0112-14-1014), the Development of Open Platform for Service of Convergence Contents].